\documentclass[twocolumn,numberedappendix,twocolappendix,iop,apj]{aastex631}
\usepackage{color,graphicx,natbib}
\usepackage{graphicx}
\usepackage{amsmath}
\usepackage{amssymb}
\usepackage{bm}
\usepackage[toc,page]{appendix}
\usepackage[T1]{fontenc}
\usepackage{xspace}
\usepackage{hyperref}
\hypersetup{
    colorlinks=True,
    citecolor=orange,
    linkcolor=black,
    }

\newcommand{\Fermi}{\textit{Fermi}}

\newcommand{\p}{\ensuremath{^{\prime}}}

\newcommand{\erg}{~\ensuremath{\rm erg}}
\newcommand{\cm}{~\ensuremath{\rm cm}}
\newcommand{\s}{~\ensuremath{\rm s}}

\newcommand{\grb}{GRB~211211A\xspace}

\shorttitle{GRB~211211A}
\shortauthors{Veres et al.}

\begin{document}
\title{Extreme Variability in a Long Duration Gamma-ray Burst Associated with a Kilonova}

\author[0000-0002-2149-9846]{P.~Veres}
\affiliation{Department of Space Science, University of Alabama in Huntsville, Huntsville, AL 35899, USA}
\affiliation{Center for Space Plasma and Aeronomic Research, University of Alabama in Huntsville, Huntsville, AL 35899, USA}

\author[0000-0001-7916-2923]{P.~N.~Bhat} 
\affiliation{Center for Space Plasma and Aeronomic Research, University of Alabama in Huntsville, Huntsville, AL 35899, USA}

\author[0000-0002-2942-3379]{E.~Burns} 
\affil{Department of Physics \& Astronomy, Louisiana State University, Baton Rouge, LA 70803, USA}
%

\author[0000-0003-0761-6388]{R.~Hamburg}
\affiliation{Universit\'e Paris-Saclay, CNRS/IN2P3, IJCLab, 91405 Orsay, France}

\author[0000-0002-0173-6453]{N.~Fraija} 
\affil{Instituto de Astronom\'ia, Universidad Nacional Aut\'onoma de M\'exico, Circuito Exterior, C.U., A. Postal 70-264, 04510, CDMX, Mexico}

\author[0000-0001-9201-4706]{D.~Kocevski} 
\affil{NASA Marshall Space Flight Center, Martin Rd. SW, Huntsville, 35808, AL, USA}

\author[0000-0003-1626-7335]{R.~Preece} 
\affiliation{Center for Space Plasma and Aeronomic Research, University of Alabama in Huntsville, Huntsville, AL 35899, USA}

\author[0000-0002-6269-0452]{S.~Poolakkil}
\affiliation{Department of Space Science, University of Alabama in Huntsville, Huntsville, AL 35899, USA}
\affiliation{Center for Space Plasma and Aeronomic Research, University of Alabama in Huntsville, Huntsville, AL 35899, USA}

\author[0000-0002-6870-4202]{N. Christensen} 
\affil{Universit\'e C\^ote d’Azur, Observatoire de la C\^ote d’Azur, CNRS, Artemis, Nice 06300, France}

\author[0000-0002-4618-1674]{M. A. Bizouard} 
\affil{Universit\'e C\^ote d’Azur, Observatoire de la C\^ote d’Azur, CNRS, Artemis, Nice 06300, France}

\author[0000-0001-5078-9044]{T. Dal Canton} 
\affil{Universit\'e Paris-Saclay, CNRS/IN2P3, IJCLab, 91405 Orsay, France}


\author[0000-0002-6657-9022]{S.~Bala}
\affiliation{Science and Technology Institute, Universities Space Research Association, Huntsville, AL 35805, USA}

\author[0000-0001-9935-8106]{E.~Bissaldi}
\affiliation{Dipartimento Interateneo di Fisica, Politecnico di Bari, Via E. Orabona 4, 70125, Bari, Italy}
\affiliation{INFN - Sezione di Bari, Via E. Orabona 4, 70125, Bari, Italy}

\author[0000-0003-2105-7711]{M.~S.~Briggs}
\affiliation{Department of Space Science, University of Alabama in Huntsville, Huntsville, AL 35899, USA}
\affiliation{Center for Space Plasma and Aeronomic Research, University of Alabama in Huntsville, Huntsville, AL 35899, USA}

\author[0009-0003-3480-8251]{W.~Cleveland}
\affiliation{Science and Technology Institute, Universities Space Research Association, Huntsville, AL 35805, USA}

\author[0000-0002-0587-7042]{A.~Goldstein}
\affiliation{Science and Technology Institute, Universities Space Research Association, Huntsville, AL 35805, USA}

\author[0000-0001-9556-7576]{B.~A.~Hristov}
\affiliation{Center for Space Plasma and Aeronomic Research, University of Alabama in Huntsville, Huntsville, AL 35899, USA}

\author[0000-0002-0468-6025]{C.~M.~Hui}
\affiliation{ST12 Astrophysics Branch, NASA Marshall Space Flight Center, Huntsville, AL 35812, USA}

\author[0000-0001-8058-9684]{S.~Lesage}
\affiliation{Department of Space Science, University of Alabama in Huntsville, Huntsville, AL 35899, USA}
\affiliation{Center for Space Plasma and Aeronomic Research, University of Alabama in Huntsville, Huntsville, AL 35899, USA}

\author[0000-0002-2531-3703]{B.~Mailyan}
\affiliation{Department of Aerospace, Physics and Space Sciences, Florida Institute of Technology, Melbourne, FL 32901, USA }

\author[0000-0002-7150-9061]{O.~J.~Roberts}
\affiliation{Science and Technology Institute, Universities Space and Research Association, 320 Sparkman Drive, Huntsville, AL 35805, USA.}

\author[0000-0002-8585-0084]{C.~A.~Wilson-Hodge}
\affiliation{ST12 Astrophysics Branch, NASA Marshall Space Flight Center, Huntsville, AL 35812, USA}

%
%
%
%
%

\begin{abstract}
The recent discovery of a kilonova from the long duration gamma-ray burst, GRB 211211A, challenges classification schemes based on temporal information alone. Gamma-ray properties of GRB 211211A reveal an extreme event, which stands out among both short and long GRBs. 
We find very short variations (few ms) in the lightcurve of \grb and estimate $\sim$ 1000 for the Lorentz factor of the outflow. We discuss the relevance of the short variations in identifying similar long GRBs resulting from compact mergers. Our findings indicate that in future gravitational wave follow-up campaigns, some long duration GRBs should be treated as possible strong  gravitational wave counterparts. 
\end{abstract}

\keywords{gamma rays: individual (211211A)}

\section{Introduction}
\label{text:intro}

Gamma-ray bursts (GRBs) are typically classified into long or short groups based on the duration of the active gamma-ray episode. Such a classification has historical origins \citep{kouveliotou93}, and the physical understanding behind this picture has matured over the following decades:
short GRBs (sGRBs) are  predominantly from binary neutron star mergers \citep{1992ApJ...392L...9D, 1992Natur.357..472U, 1994MNRAS.270..480T, Goldstein+17170817a,Abbott2017aGWGRB} or possibly from black hole - neutron star mergers \citep{1992ApJ...395L..83N}, while long GRBs (lGRBs) originate from the core collapse of massive stars \citep{1993ApJ...405..273W,1998ApJ...494L..45P, Macfadyen+99col,Woosley2006ARA&A}. There is an overlap between the duration distributions of short and long classes. For this reason, the classification based on the burst duration is complemented by rudimentary spectral information, the hardness ratio (HR), available for all GRBs. Classifications based on two parameters provide better separation between the classes. On average, sGRBs are harder, while lGRBs are softer \citep{Paciesas+99cat,Bhat+16cat, vonKienlin+20GBM10yrcat}. In some cases even two parameters are not sufficient to derive a reliable classification and further observations are needed to hone in on the physical origin of the GRBs (see e.g. the Type I/II classification scheme by \citet{Zhang+09typeI} and refer to \cite{kann2011} for a discussion of controversial scenarios).
LGRBs also include Ultra-long GRBs (ULGRBs) with a duration longer than thousands of seconds  \citep{Gendre2013, levan2014, Piro2014, Greiner2015Natur,Kann2018A&A}, and lGRBs associated with Supernovae (SNe) Ic \citep{hjorth03}.  Additionally giant flares from extragalactic magnetars \citep{Roberts+21MGF} can masquerade as sGRBs at a rate of approximately one event per year \citep{burns2021identification} further complicating the picture.

\begin{figure*}
\includegraphics[width=1.0\textwidth]{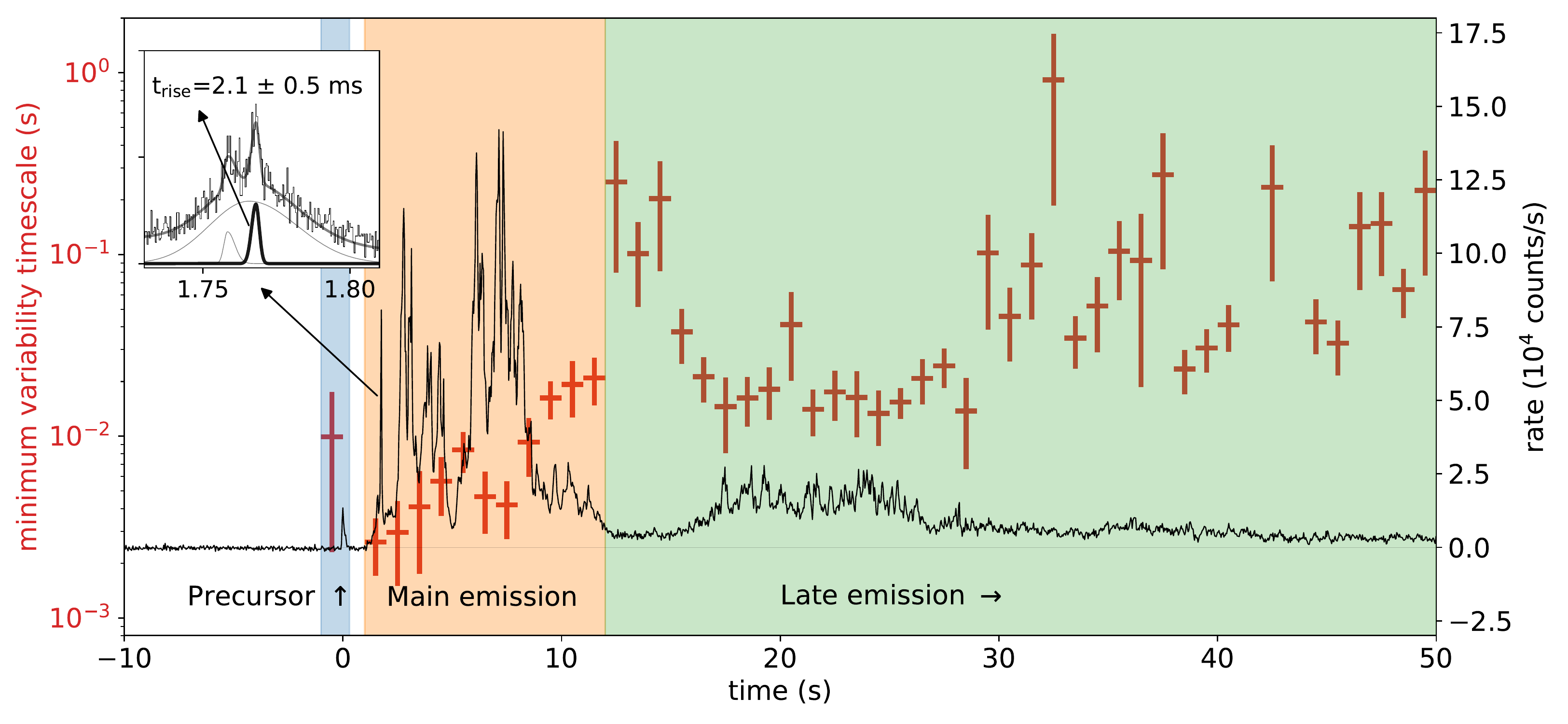}
\caption{Lightcurve of \grb (black) and the corresponding minimum variability timescale (red). Inset: Zoomed lightcurve around the time of the shortest variation. Individual pulse models are indicated, with the shortest rise time highlighted in bold.}
\label{fig:lc0}
\end{figure*}

At first glance, \grb is a bright, but otherwise typical lGRB suggesting a collapsar origin, based on the gamma-ray properties. However, \citet{Rastinejad+22grb211211a,Troja+22grb211211a} report a possible kilonova counterpart to \grb suggesting a compact merger origin, at odds with the gamma-ray classification.  \grb  represents one of the clearest breaks with the usual short/long classification. 
Some sGRB pulses are followed by a longer,  extended gamma-ray emission, without associated supernovae  \citep{Gehrels+06strangegrb}. It is possible that \grb also belongs to this class.

One of the most intriguing features of GRBs is the short variations in their lightcurves. The observed variability could originate from the variations in the central engine with contributions from the jet interacting with the progenitor  \citep{Sari+97intext,Morsony+10variab}. This variability is imprinted on the emission processes \citep[e.g. internal shocks][]{Sari+97variab}, or alternatively can be ascribed to intrinsic variations in the emitting volume \citep[e.g. by turbulence][]{Narayan+09turb}. Typical variations can be as short as 10 ms, with a handful of examples of sub 10 ms variability. The most extreme case for a GRB is a $\sim$200~$\mu$s variation \citep{Bhat+92shortvar}.  On average, sGRBs have shorter variability than lGRBs \citep{Bhat+12mvt,Golkhou+15tvar}. The variability timescale constrains the size of the emitting region based on causality arguments \citep{rybicki79}. 

Here, we place \grb in the context of {\Fermi} Gamma-ray Burst Monitor (GBM) GRBs and report on the implications for future gravitational wave (GW) or kilonova searches associated with GRBs.
We provide a detailed analysis of the gamma-ray lightcurve  highlighting the extreme variability 
and discuss this the possible association of this GRB with the class of short GRBs with extended emission. 

We present gamma-ray observations of \grb in Section \ref{sec:data}, focusing on the minimum variability timescale (Section \ref{sec:mvt}). We present \grb as a GRB with extended emission, and provide physical parameters of the outflow in Section \ref{sec:results}. We end with discussing our results in Section \ref{sec:discussion}.
We use the $Q_x=Q/10^x$ convention in c.g.s. units for quantity $Q$ and refer to physical constants using their common notations.

\section{Data analysis}
\label{sec:data}
\grb\citep{2021GCN.31210....1M} triggered {\it Fermi}-GBM \citep{Meegan2009} at 13:09:59.65 UT on December 11th, 2021 ($T_{0}$). It showed significant emission in all 12 of GBM's NaI and both the BGO detectors, up to an energy $\sim$ 20 MeV. At trigger time the location of the GRB was outside of the LAT field of view, however \citet{Mei+22211211aGeV} reported detection of photons in the GeV range at $\sim T_0+10^4 \s$. Swift-BAT \citep{2021GCN.31202....1D}, CALET \citep{2021GCN.31226....1T}, INTEGRAL-SPI/ACS \citep{2021GCN.31230....1M} and Insight-HXMT \citep{2021GCN.31236....1Z} also detected \grb.

The duration $T_{90}=34.3\pm0.6 \s$ calculated as the central $90^{\rm th}$ percentile of the cumulative energy flux in the 50-300 keV range using the  {\tt RMfit}\footnote{\url{https://fermi.gsfc.nasa.gov/ssc/data/analysis/rmfit}} software. 
The hardness ratio over the T$_{90}$ duration, defined as the ratio of fluxes between  50-300 keV  and   10-50 keV energy ranges is $HR= 0.850 \pm 0.015$. These two parameters place \grb on the duration-hardness plane with high probability in the long population \citep{Rastinejad+22grb211211a,Bhat+16cat,vonKienlin+20GBM10yrcat,Rouco+21180418}. For spectral analysis we use NaI detectors {\tt n2} and {\tt na} and BGO detector {\tt b0}. For temporal analysis we can use additional NaI detectors with significant flux: {\tt n0}, {\tt n1}, {\tt n2}, {\tt n5}, {\tt n9}, {\tt na} and {\tt nb}.

\subsection{Lightcurve}
\label{sec:lc}

Morphologically, the lightcurve can be separated into three parts. This GRB starts with a brief standalone pulse, a precursor, lasting about 0.2 s. Interestingly \citet{Xiao+22211211aperiod}, reported tentative quasi-periodic oscillations for this pulse. The second part is the brightest and we refer to it as {\it the main emission} episode. It starts at T$_{0}$+1 s and lasts until T$_{0}$+13 s. It consists of a large number of short peaks. 
The third part starts around $T_{0}$+13 s, and we refer to it as {\it late emission}. It contains less variability than the main emission, it can be detected until about $T_{0}$+70 s  and it fades smoothly into the background (see Figure \ref{fig:lc0}).

Taken by itself, with duration of $\approx 10 \s$, even the {\it main emission} episode would be classified as a lGRB. It is significantly longer than the $\approx 4-5$ s limit separating the short and long classes of GBM \citep{vonKienlin+20GBM10yrcat}. This limit represents the duration of equal probability between the long and short classes when we model the duration distribution using two log-normal components (see e.g. section \ref{sec:tail} and the associated figure).

\subsection{Spectrum}

To compare \grb with other GRBs, we perform a  spectral analysis of the brightest peak and the time integrated emission. We note that the time integrated analysis, with fluence $F=(5.1\pm0.1)\times 10^{-4}\erg \cm^{-2}$ does not capture the evolving trends observed in this burst by e.g. \citet{Gompertz+22grb211211a} but it is suitable to determine the gamma-ray energetics. The peak flux, commonly reported on 64 ms (sGRBs) and 1.024 s (lGRBs) timescales is $P_{64{\rm ms}}=(1.49\pm0.02)\times 10^{-4}\erg \cm^{-2} \s^{-1}$ and $P_{1{\rm s}}=(8.10\pm0.04)\times 10^{-5}\erg \cm^{-2} \s^{-1}$, respectively.
Both the time integrated and the peak spectra are best fit by Band functions \citep{Band+93} with parameters presented in Table \ref{tab:params}.

The redshift reported for the host galaxy is z=0.076 \citep{Malesani+21211211aredshift}, corresponding to a luminosity distance of $D_L$=346 Mpc (using $\Omega_m=1-\Omega_\Lambda=0.315$ and $H_0=67.4 ~{\rm km} \s^{-1} ~{\rm Mpc}^{-1}$ \citep{2020A&A...641A...6P}). The isotropic-equivalent gamma-ray energy of \grb calculated in the 1-10,000 keV range is
$E_{\rm iso}\approx 1.3\times 10^{52}\erg$, the peak luminosity, calculated on a 64 ms and 1.024 s timescale is   $L_{\rm iso,64 ms}\approx5.9\times 10^{51} \erg \s^{-1}$  and $L_{\rm iso,1 s}\approx2.3\times 10^{51} \erg \s^{-1}$ respectively (see Table \ref{tab:params}).

\subsection{\grb in context of other GRBs}

\grb has higher energy fluence  (units of erg cm$^{-2}$)  than all but 4 GRBs in the GBM catalog \citep{vonKienlin+20GBM10yrcat} (GRBs 130427A, 161625B, 171010A and 160821A), corresponding to 99.9${^{\rm th}}$ percentile among Fermi GBM GRBs \citep{Poolakkil+21GBMspcat}. 
The peak energy flux  (units of erg cm$^{-2}$ s$^{-1}$) of \grb calculated for the brightest 64 ms is brighter than all short GRBs in the catalog. The 1.024 s peak energy flux of \grb is brighter than all but two long GRBs prior to \grb (GRB 130427A \citep{Preece+14130427agbm}, and GRB 131014A \citep{Guiriec+15131014ashortbright}).

During the writing of this paper, \Fermi-GBM detected GRB~221009A \citep{Lesage+23221009a} and GRB~230307A \citep{Dalessi+23230307a} with peak fluxes and fluence larger than \grb. While GRB~221009A is clearly not from a compact binary merger \citep{Fulton+23221009asupernova}, GRB~230307A does bear some resemblance to \grb (see section \ref{sec:other}).

The peak energy (E$_{\rm peak}$) measured for the brightest 1 s (1030 keV) is 94.9 percentile among lGRBs and 84.9 percentile among sGRBs.

We conclude that \grb is at the bright end of peak flux and fluence distributions among both short and the long classes, making it an exceptional GRB in the Fermi GBM sample.

\section{Minimum variability timescale}
\label{sec:mvt}

The minimum variability timescale (MVT) of a GRB lightcurve represents the shortest timescales, at which coherent changes can be identified. In practice it coincides with the typical timescale (e.g. the rise time) of the shortest pulse in the lightcurve.
There are multiple mathematical methods in the literature to derive the MVT. Here we use the methods of
\citet{Bhat+12mvt,Bhat13tvar} and \citet{Golkhou+15tvar}, but see also \citet{MacLachlan+13tvar}.

We binned our lightcurve to $100~\mu$s and searched for the shortest coherent variations.
The variability using the method of \citet{Bhat+12mvt}  is $\Delta t_{\rm var}= 2.6\pm 0.9$ ms.
The \cite{Golkhou+15variability} method gives a variability of $\Delta t_{\rm var}=2.5\pm0.8$ ms. The two methods are independent, and they give consistent MVT values, strengthening the confidence that this is indeed the minimum variability timescale of this burst.

In  addition, we identify a pulse with rise time of $\approx 2$~ ms in Figure \ref{fig:lc0} (inset) that determines the MVT: we fit the high resolution (400 $\mu$s) lightcurve in the range 1.73 to 1.81 s, (region of the lightcurve where the variability time is the shortest) with the pulse model of \citet{Norris+05pulse}, using 3 pulses plus a long term emission modeled as a first degree polynomial. The rise time of the shortest pulse is consistent with minimum variability timescale, as expected \citep{Bhat+12mvt}.
This $\sim2$~ms timescale is significantly lower than the 16 ms variability reported by \citet{Yang+22211211a} and the 10 ms reported by \citet{Xiao+22211211aperiod}

We also performed a time resolved variability analysis. We calculate the MVT in each 1 s bin (Figure \ref{fig:lc0}) and find that the separation of the lightcurve into {\it main} and {\it late} emission is also  reflected in the evolution of the variability timescale: the main emission episode has a clearly shorter variability than the late emission.

\subsection{Long duration GRBs with short MVT}
\begin{figure}
\includegraphics[width=1.0\columnwidth]{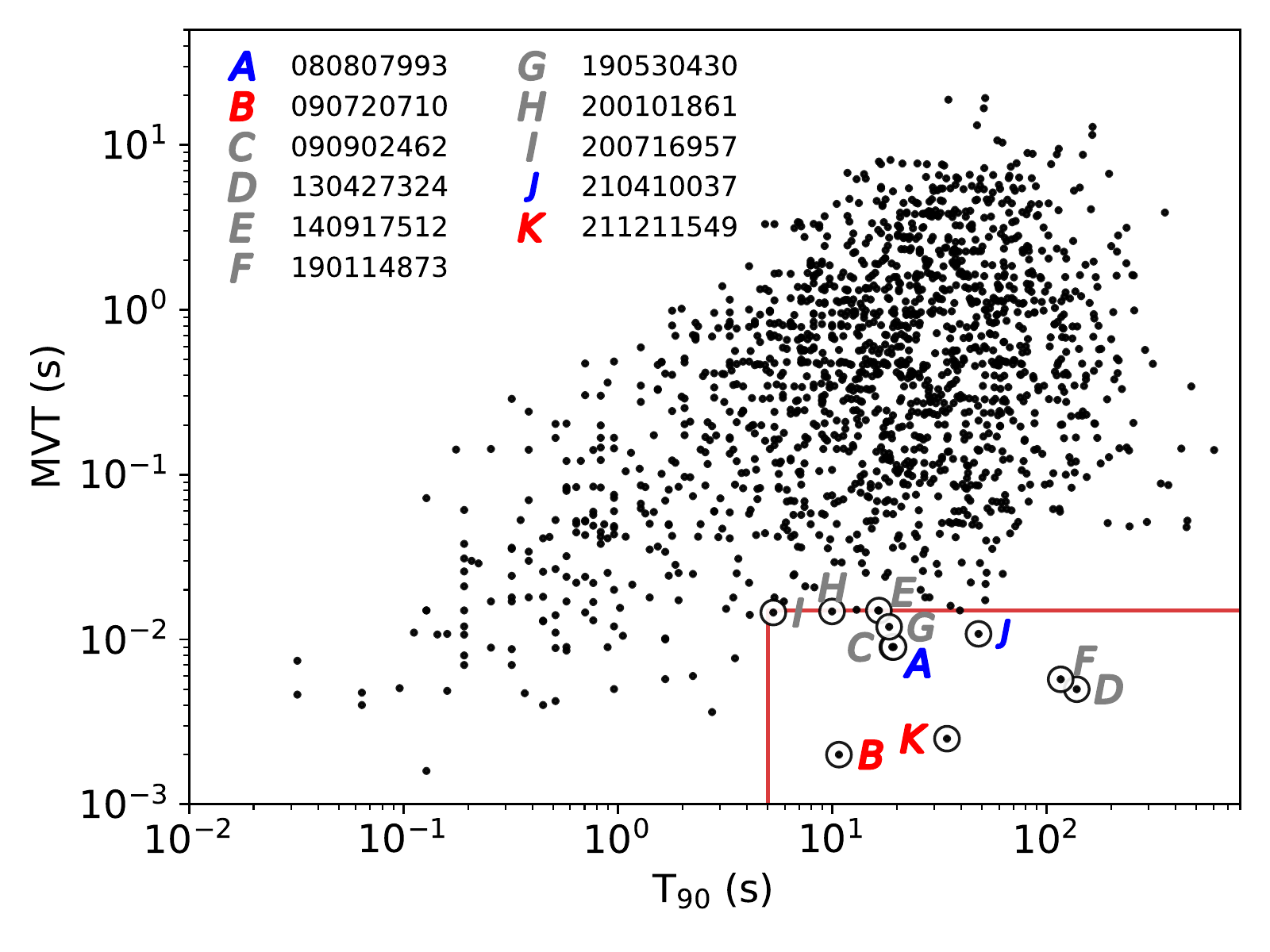}

\caption{MVT values as a function of T$_{90}$ for all Fermi-GBM GRBs with well measured T$_{90}$ and MVT. We highlight GRBs with T$_{90}>5\s$ and MVT$ <15$~ms (red line) by displaying the GBM trigger number in the legend. \grb and GRB 090720B are indicated by red letters ({\bf B}, {\bf K}), blue letters ({\bf A}, {\bf J}) mark two possibly similar GRBs, but with higher MVT, other selected GRBs have gray letters.}
\label{fig:manyMVT}
\end{figure}

\grb, with MVT$\sim 2.5$~ms is a clear outlier in the distribution of the MVTs presented in \citet{Golkhou+15tvar} (see Figure \ref{fig:manyMVT}, where we plot MVT values that have an uncertainty smaller than the value itself). We have searched for other GRBs that have long duration (T$_{90}\gg 2\s$), and short variability. The sample of \citet{Golkhou+15tvar} contained MVT values only until 2012. We extended their sample with bursts up to 2022, by additional 2124 \Fermi-GBM GRBs.

Because the MVT calculation depends on multiple input parameters, that can affect the final value (e.g. detector selection, background, foreground interval, method, etc.),  we  allow a limit of MVT$<$15 ms and $T_{90}>5\s$ in searching for GRB similar to \grb (see Figure \ref{fig:manyMVT}).

Applying the above limit, we found 10 potentially interesting GRBs in our sample (see Figure \ref{fig:manyMVT}). Many of the selected bursts are known, bright GRBs with associated supernovae (e.g. GRB~130427A \citep{Preece+14130427agbm} and GRB 190114C \citep{Ajello+20190114c}), or their lightcurve does not resemble that of \grb. We inspected each GRB lightcurve visually, looking for similar lightcurve morphology to \grb, namely an initial bright, variable phase followed by a longer, less luminous emission episode. 
After visual inspection of the candidates, we find 3 additional cases with similar lightcurves as \grb. Among these three, only GRB 090720B has comparable variability timescale ($\sim$ 2 ms), while other similar GRBs: 210410A and 080807 have $\sim$10 ms variability. We thus conclude that bursts with long duration and short variability are rare, especially those GRBs that exhibit a short pulse followed by softer, extended emission.
We further conclude that \grb lies at the extreme low end of the variability timescale distribution of Fermi-GRBs.

\begin{table}[]
    \centering
    \begin{tabular}{c|c|c}
 GRB& T$_{90}$(s)& MVT(s)\\ \hline
120712571& $22.528\pm 5.431$ & $0.6646\pm 0.1799$ \\   
120716577& $24.960\pm 3.958$ & $7.7225\pm 1.8755$\\
120728934& $32.768\pm 2.429$ & $1.5558\pm 0.6006$\\
120805706& $1.856 \pm 1.296$ & $0.6957\pm 0.1710$\\
120806007& $26.624\pm 1.557$ & $0.2707\pm 0.1133$\\
120811014& $0.448 \pm 0.091$ & $0.0182\pm 0.0066$\\
\hline
    \end{tabular}
    \caption{Table of MVT values for GBM GRBs. The table contains GRBs starting from the end of GRBs covered in \citet{Golkhou+15tvar}. The table is available in its entirety in machine-readable form.}
    \label{tab:allmvt}
\end{table}

\begin{deluxetable*}{ccccccccc}
    \label{tab:params}
 \tabletypesize{\footnotesize}
 \tablecolumns{8}
 \tablecaption{Spectral parameters of \grb}
 \tablehead{   
   \colhead{Time (s)} &
   {E$_{\rm peak}$} &
   {$\alpha$} &
   {$\beta$} &
   {Energy Flux }&
   {Fluence }&
   {L$_{\rm iso}$}& 
   {E$_{\rm iso}$} \\
    \colhead{T-$T_{0}$ (s)} &
   {keV} &
   {} &
   {} &
   {$10^{-5} \erg \cm^{-2} \s^{-1}$ }&
   {$10^{-5} \erg \cm^{-2}$ }&
   {$10^{51} \erg \s^{-1}$}& 
   {$10^{52} \erg$} 
   }
\startdata
0-52.2       &  $545^{+12.3}_{-10.8} $  &  $-1.180 ^{+0.005}_{-0.006}$ & $ -2.13^{+0.02}_{-0.019}$ & -               
& $50.7\pm0.1$  & - &   $ 1.25\pm0.003$    \\
7.104-7.168  &  $1737^{+114}_{-121} $   &  $-0.878 ^{+0.019}_{-0.020}$ & $ -2.90^{+0.26}_{-0.19}$  &  $14.9\pm0.2 $  
& - &  $5.89\pm0.09$  & -   \\
7.168-8.192  &  $1033^{+27}_{-31} $     &  $-0.941 ^{+0.008}_{-0.008}$ & $ -2.72^{+0.07}_{-0.06}$  &  $8.10\pm0.04$  
&- &  $ 2.33\pm0.01$ & -    \\
\enddata
\tablecomments{Spectral parameters for \grb fitting a Band function and using the standard time intervals: the entire GRB, brightest 1024 ms and 64 ms. The flux and fluence are reported in the 10-1,000 keV range. L$_{\rm iso}$ and E$_{\rm iso}$ are reported in the 1-10,000 keV (observer) range.}
\end{deluxetable*}

\section{Results}
\label{sec:results}
\subsection{\grb as a short GRB with extended emission}
 
GRB 060614 \citep{Gehrels+06strangegrb}, a nearby long event with duration in excess of 100 s has  no supernova detection to deep limits, indicating a possible merger origin. Its lightcurve morphology, a short pulse, followed by extended emission established a new category of GRBs (sGRB-EE). GRB 060614 was detected by Swift-BAT. For this GRB we derive an MVT value of $\Delta t_{\rm var} = 36  \pm   4$~ms.
\cite{Rastinejad+22grb211211a,Gompertz+22grb211211a,Xiao+22211211aperiod, Troja+22grb211211a} find that \grb has broadly consistent properties with other sGRB-EE GRBs. We also find that \grb has consistent features with sGRB-EE. The time-resolved MVT (Figure \ref{fig:lc0}) also clearly delineates the sGRB and the extended emission.

\citet{Kaneko+15ee} considered a sample of sGRB-EE in the \Fermi-GBM sample. We extend their list (see Table \ref{tab:ee}) and investigate the variability timescale of sGRB-EE (Burns et al. in preparation). We find that \grb has shorter variability timescale than all the GBM sGRB-EE. This means that the MVT of \grb is extreme also among the short GRBs with extended emission (including the archetypal GRB 060614). 
We also note that the MVT of \grb is close to the short end of even the short duration GRBs (see Figure \ref{fig:manyMVT}).

\begin{table}[]
    \centering
    \begin{tabular}{c|c|c}
Trigger number & MVT (ms) & T$_{90}$ (s)\\ \hline
081110601 & $291 \pm 11$   & $11.8 \pm 2.6$\\
090227772 & $\sim 5 $      & $1.28 \pm 1.03$\\
090510016 & $5 \pm 1$      & $0.96\pm0.14$\\
090831317 & $15 \pm 4$     & $39.4\pm0.6$\\
100916779 & $88 \pm 8$     & $12.8\pm2.1$\\
111221739 & $18 \pm 6$     & $27.1\pm7.2$\\
140819160 & $43 \pm 20$    & $6.7\pm3.7$\\
170728961 & $54 \pm 15$    & $46.3\pm0.8$\\
180618030 & $\sim 7 $      & $3.7\pm0.6$\\
190308923 & $\sim 419 $    & $45.6\pm2.83$\\
200219317 & $\sim 55 $     & $1.15\pm1.03$\\
200313456 & $\sim 284 $    & $5.2\pm4.4$\\
201104001 & $25 \pm 8$     & $52.5\pm7.4$\\
    \end{tabular}
    \caption{Sample of short GRBs with extended emission. Values without errors represent cases where the signal-to-noise wasn't sufficient to determine an error.}
    \label{tab:ee}
\end{table}

\subsection{Possible afterglow origin of the late emission}
As noted in the previous section, \grb fits into the category of short GRB with extended emission \citep{Gehrels+06strangegrb,Norris+10EE}. The {\it main emission} plays the role of the short GRB, and the {\it late emission} episode corresponds to the extended emission.

The origin of the extended emission is unclear, sometimes it is associated with late energy injection into the GRB \citep[e.g.][]{Bucciantini+12sgrbEE}.
In many cases, the extended emission has appreciable variability and for this reason its association with afterglow emission is generally disfavored \citep{norris06}. 
Based on the fact that the {\it late emission}  ($T_{0}$+12 to $T_{0}$+70 s in Figure \ref{fig:lc0}) has longer variability timescale  than the {\it main emission}, we explore the afterglow origin for the {\it late emission}. In this scenario the late emission is emitted by the shocked circumstellar medium as it slows down the shells that were responsible for the prompt emission. Detecting the afterglow  in the $\gamma$-ray regime has been reported before \citep[e.g.][]{Giblin+99afterglow,connaughton02,Ajello+20190114c} and it is common for bright GRBs.

\begin{figure}
\includegraphics[width=1.0\columnwidth]{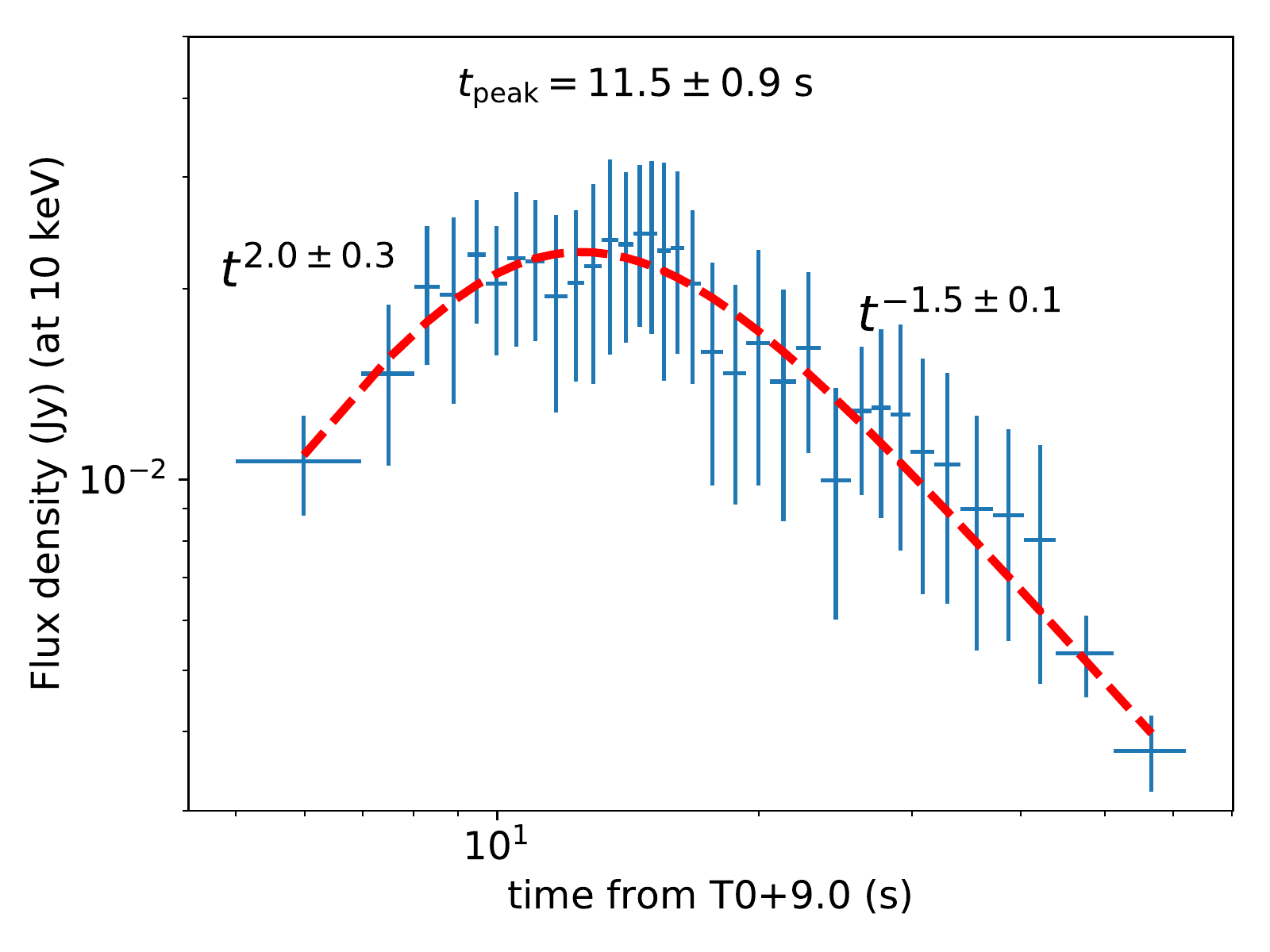}
\caption{Flux density evolution of the late emission, fit with a smoothly broken power law. The time axis is shifted by 9 s with respect to the trigger time to match the end of the main episode.}
\label{fig:decel}
\end{figure}

Early afterglow lightcurves, especially in X-ray and GeV sometimes show a rising phase, peak and a decay representing the onset of the afterglow. Here the peak marks the deceleration time. 
We binned the GBM lightcuve from 13 to 70 s after the trigger into bins with signal to noise ratio of 60. We fit each spectrum with a Comptonized function (power law with exponential utoff) and calculate the flux density at 10 keV (see Figure \ref{fig:decel}). The flux evolution shows a clear peak. We fit the flux density curve with a smoothly broken power law function, $f(t)\propto\left[(t/t_{\rm peak})^{s\alpha_{\rm rise}} + (t/t_{\rm peak})^{s\alpha_{\rm decay}}\right]^{-1/s}$, where $s$ is the smoothness parameter, fixed here to 1. We find the index of the rising phase $\alpha_{\rm rise}=2.0\pm0.3$ consistent with the expectation of $\alpha_{\rm rise}=2$ if it originates from the forward shock before deceleration \citep{Sari+99optflash}. We note that $\alpha_{\rm rise}$ is sensitive to the choice of the zero point. Here we choose T$_{0}$ + 9 s, because this is  the approximate end time of the highly variable, main emission episode. The peak of the flux occurs at $20.5\pm0.9 \s$ after the trigger time.
The temporal decay index after the peak is $\alpha_{\rm decay}=-1.5\pm0.1$, which in the forward shock scenario  corresponds to $1/2 -3p/4$, where $p$ is the index of shocked electron population's the power law distribution. In our case we get $ p= 2.7 \pm 0.1$, which is consistent with the values that are commonly found for afterglows \citep[e.g.][]{panaitescu01}. 

We thus conclude that the afterglow interpretation is possible at least in some cases of GRBs with extended emission.

\subsection{Lorentz factor constraints}
\label{sec:Lorentz}
The Lorentz factor of the outflow is a basic ingredient of the physical picture.  
We can provide a lower limit on the Lorentz factor if the highly variable prompt gamma-ray lightcurve is produced by internal shocks \citep{Rees+94is}. In this scenario the emission radius ($R_{\rm IS}$) has to be above the photosphere ($R_{\rm ph}$) where the optical depth is unity. The internal shock radius is $R_{\rm IS}\approx 2c\Gamma^2 \Delta t_{\rm var}$, while the photosphere radius is $R_{\rm ph}=L\sigma_T/8\pi m_p c^3\Gamma^3$, assuming the photosphere occurs in the coasting phase (i.e. the jet does not accelerate any more, $\Gamma(R)=$constant).
From $R_{\rm IS}\gtrsim R_{\rm ph}$, we have
\begin{eqnarray}
    \Gamma\gtrsim170 \left(\frac{L_{\gamma,51.8}}{\eta_{\gamma,-0.7}} \right)^{1/5}\left( \frac{\Delta t_{\rm var}}{2.5~{\rm ms}}\right)^{-1/5}.
\end{eqnarray}
Meaningful Lorentz factor constraint using this method is only possible for GRBs with high luminosity and short variability, uniquely relevant for \grb.

We can calculate the bulk Lorentz factor by identifying the peak of Figure \ref{fig:decel} with the onset of the afterglow or the deceleration time. The deceleration radius corresponding to the deceleration time ($t_{\rm dec}$) marks the distance from the central engine where the relativistic outflow has plowed up interstellar matter that is a fraction $\sim1/\Gamma$ of the jet mass. 
Using the expression of $t_{\rm dec}$, the Lorentz factor evolving a constant-density medium will be:
\begin{eqnarray}
    \Gamma &=&\left( \frac{3E_{\rm k}}{4\pi m_p c^5 n}\right)^{1/8} t_{\rm dec}^{-3/8}\\
    &\approx& 1200 \left
    (\frac{E_{\gamma,52.1}}{{n_{-4}\eta_{\gamma,-0.7}}}\right)^{1/8}\left(\frac{t_{\rm peak}}{11.5\s}\right)^{-3/8}\,.
\end{eqnarray}

Here we chose a gamma-ray efficiency of  $\eta_\gamma=E_\gamma/E_{\rm k}
= 7.8\%$, where $E_k$ is the kinetic isotropic equivalent energy, and  $n=10^{-4}\cm^{-3}$ the constant interstellar number density, as scaling values, from \citet{Mei+22211211aGeV}. If we conservatively  measure the peak from the trigger time, instead of the $9 \s$ shift that we introduced, the Lorentz factor becomes $\Gamma\approx 980$.

\citet{Sonbas+15LorentzMTS} present a correlation between variability timescale and Lorentz factor based on a compilation of Lorentz factor estimates. The relationship they find is $t_{\rm var}(\Gamma)\propto \Gamma^{-4}$ for $\Gamma\gtrsim 200$ and $t_{\rm var}\approx$ constant otherwise. Inverting the correlation, and substituting $\Delta t_{\rm var}=2.5$ ms, we get $\Gamma\approx900$, which is consistent with the above estimates. Furthermore, the Lorentz factor estimates from different methods are consistent with the best estimate by \citet{Mei+22211211aGeV} of $\log\Gamma\approx 3.1^{+0.9}_{-0.6}$. In the internal shock scenario the emission radius of the gamma-rays will be:
\begin{equation}
    R_{\gamma}\approx2 \Gamma^2 c\Delta t_{\rm var}= 1.5\times 10^{14}~ \Gamma_3^2\left( \frac{\Delta t_{\rm var}}{2.5~{\rm ms}}\right) \cm.
\end{equation}

\subsection{Event rate: the tail of the merger distribution}
\label{sec:tail}
\begin{figure}
\includegraphics[width=1.0\columnwidth]{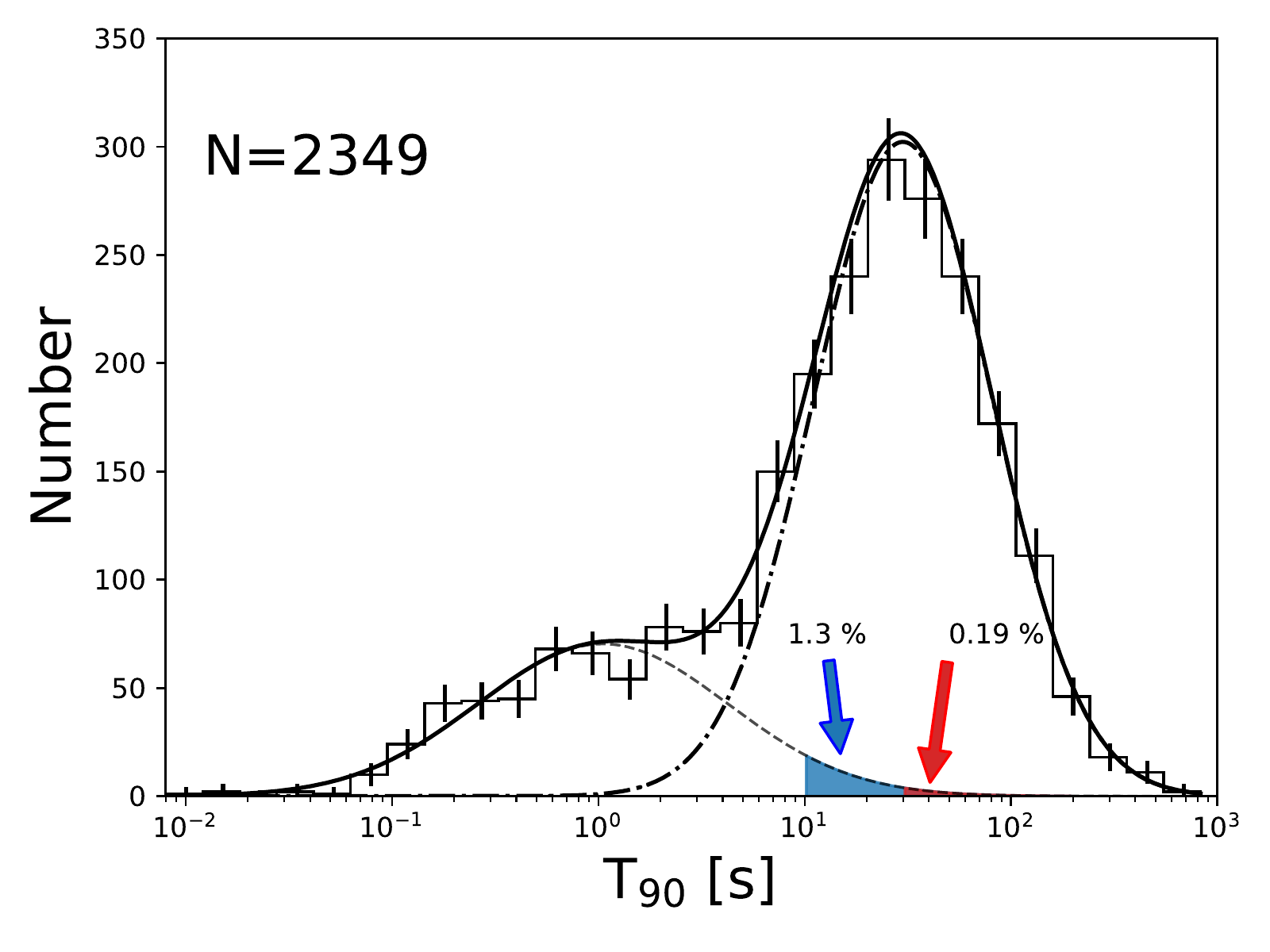}
\caption{T$_{90}$ distribution of 10 years of GBM GRBs \citep{vonKienlin+20GBM10yrcat}. Fractions of short GRBs with $T_{90}>10\s$ and $T_{90}>30\s$ are indicated.}
\label{fig:t90_tail}
\end{figure}

The 10 year GBM GRB catalog \citep{vonKienlin+20GBM10yrcat} contains in excess of 2300 GRBs with duration measurements. The distribution of the T$_{90}$ durations is modeled as the sum of two log-normal functions. The exact reason why the T$_{90}$ distribution would follow a log-normal distribution is unclear \citep[for a possible explanation see][]{Ioka+02lognormal}, and in reality the distributions could be asymmetrical \citep{Tarnopolski19skewed_distrib}. Because the two component model provides a good fit to the distribution,
we will consider this description to calculate the rate for mergers masquerading as lGRBs
(see Figure \ref{fig:t90_tail}). 

We integrate the short model from 10 s, broadly corresponding to the duration of the {\it main emission} and the actual $T_{90}\sim 10 \s$ of the similarly short MVT GRB 090720B (Section \ref{sec:other}). We also integrate the short model component for $T_{90}>30 \s$ corresponding roughly to the  $T_{90}$ of \grb. We find about 1.3 \% (3 per year) of short \Fermi-GBM GRBs will have $T_{90}>$10 s and about 0.19 \% (0.4 per year) of short GRBs will have $T_{90}>$30 s.

\subsection{Other examples of long duration and short MVT}
\label{sec:other}
\begin{figure}
\includegraphics[width=1.0\columnwidth]{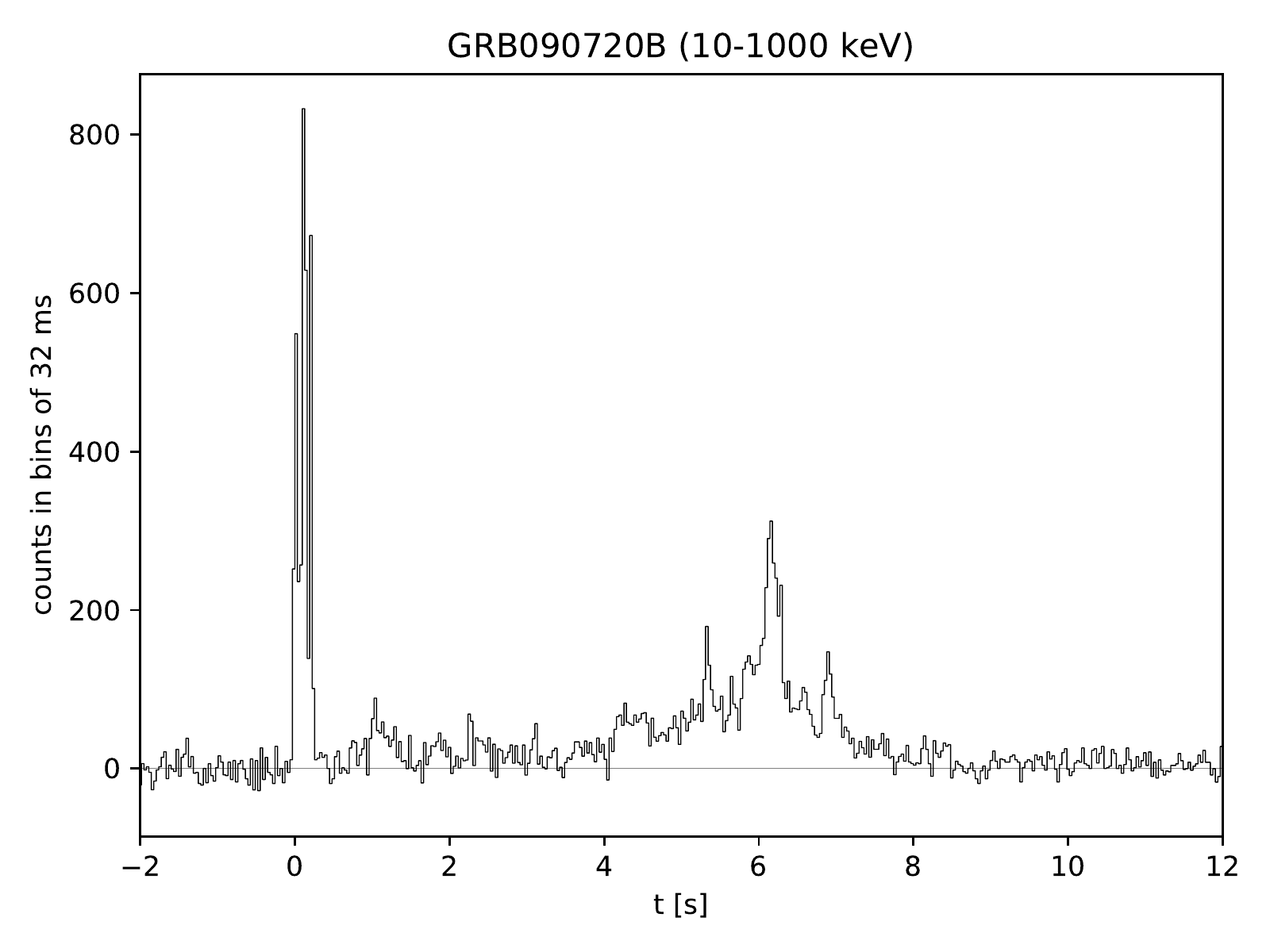}
\caption{Lightcurve of GRB~090720B with similar short pulse + extended emission structure. GRB~090720B has similarly short ($\sim 2$~ms) variability timescale as \grb.}
\label{fig:ex1}
\end{figure}

GRB~090720B (GBM trigger 090720710, \cite{2009GCN..9698....1B}) was a bright GRB, with similar properties as \grb: an initial bright, highly variable set of overlapping pulses, followed by weaker, less variable emission (Figure \ref{fig:ex1}). We cross-check the short variability measurement by \citet{Golkhou+15tvar}, $\Delta t=2\pm1$~ms, and consistently find $\Delta t=2.6\pm   0.9$ ms with the method of \citet{Bhat+12mvt}. The duration of this GRB is  $T_{90}=10.8\pm 1.1 \s$. Time resolved variability similarly shows a shorter timescale in the main emission compared to the longer lasting episode. GRB~090720B has no redshift measurement, however it is detected by \Fermi-LAT \citep{2012MNRAS.421L..14R,Ajello+19LATcat2}.

Two further examples with  larger variability timescale  ($\sim$ 10 ms) are GRB 080807 \citep{vonKienlin+20GBM10yrcat} and 210410A \citep{2021GCN.29788....1W} with $T_{90}$ of $19.1 \pm 0.2\s$ and $48.1 \pm 2.8\s$ respectively. Neither have a redshift measurement.
Except for GRB 080807, the other 3 selected GRBs have been detected at high energy by LAT. GRB 080807 occurred in a unfavorable geometry for LAT.

GRB 230307A is a recent bright GRB that has tentatively similar properties as \grb. Despite its long duration ($T_{\rm 90}\approx 35\s$, \citet{Dalessi+23230307a}) it shows  short variations like \grb and has tentative kilonova signature \citep{Levan+23230307aKN}. Because of its extreme brightness the prompt measurement suffers from instrumental effects \citep{Dalessi+23BTI}. A detailed study of GRB 230307A is left for a forthcoming paper.

\subsection{Implications for searches for GW counterparts}
Having characterized the gamma-ray emission of \grb, there still remains an intriguing question. How can a merger event give rise to a GRB that has a duration well in excess of the historical 2 s limit  between the short and long classes? Even if we place \grb at a larger distance where only the main emission episode is detectable, its duration will be $\sim$ 10 s and it will be classified as a likely long GRB. This suggests that GRBs with duration $\gtrsim$ 5 s  can possibly originate from compact binary mergers and be gravitational wave counterparts. Follow-up decisions should consider this fact.

If a sub-class of long GRBs corresponds to compact binary mergers as their source then this will have implications for the gravitational-wave signal search strategy of LIGO-Virgo-KAGRA (LVK). Currently LVK search for gravitational waves in coincidence with GRBs detected by the Fermi and Swift satellites~\citep{LIGOScientific:2020lst,LIGOScientific:2021iyk}. In these searches, GRBs are classified as short if $T_{90} < 2$ s, long if $T_{90} > 4$ s, or ambiguous for all the other cases. The times coincident with GRBs classified as short or ambiguous are searched for gravitational-wave signals from compact binary mergers using a coherent matched filter analysis, PyGRB~\citep{Harry:2010fr,Williamson:2014wma}. The merger time is assumed to fall within a $[-5,1]$ s window, where 0 corresponds to the GRB trigger time.

LVK also use an excess power analysis to search for generic transient signals associated with all GRBs, namely X-Pipeline~\citep{Sutton:2009gi,Was:2012zq}. The search window for gravitational-wave transients begins 600 s before the GRB trigger time, and stops 60 s after trigger time; if $T_{90} > 60$ s then the end of the search window is $T_{90}$.

During observing run O3, times coincident with 49 GRBs were examined targeting compact binary merger signals. The times coincident with 191 GRBs were examined with the generic search pipeline~\citep{LIGOScientific:2020lst,LIGOScientific:2021iyk}. If only the GRB $T_{90}$ is considered, as is presently the case for these LVK analyses, then including GRBs such as GRB 211211A will require a significant broadening of the ``ambiguous'' class, considerably increasing the number of GRBs that will have to be analyzed with the compact binary merger pipeline. This is not an impossible challenge, but would require significantly more human and computing resources, and would increase the chance of a false alarm from the significantly larger sample of GRBs that are not originated by compact binary mergers.

The results of this study motivate the design of a more reliable GRB classification scheme that includes the MVT in addition to $T_{90}$. 
The observation of a kilonova should obviously also be incorporated into this improved classification.

\section{Discussion and Conclusion}
\label{sec:discussion}

The observation of \grb represents one of the clearest examples that defy the duration based GRB classification scheme. 
We analyzed the gamma-ray properties of \grb in context of the \Fermi-GBM GRB population. We found that \grb is one of the brightest GRBs among both the merger and collapsar population.

We found indications that the extended emission can be modeled as early afterglow in the gamma-rays, and that leads to an estimate of the Lorentz factor. We calculated the variability timescale with different methods and conclusively found one of the shortest variation among long duration GRBs.  
The short variability has implications on the emission mechanisms and lets us determine the physical parameters of the source. We found the Lorentz factor consistent with $\Gamma\approx1000$, which puts it among the highest inferred values. 
$\Gamma\approx1000$ agrees well with the value reported by \citet{Mei+22211211aGeV} and it is larger than the values ($\approx 100$) derived by \citet{Gompertz+22grb211211a,Rastinejad+22grb211211a}.

We estimated the fraction of short GRBs based on the best fit to the duration distribution. Even though the extrapolation is uncertain, we can conclude that $\sim 3.1$ GRBs per year with merger origin will have $T_{90}>10 \s$, and $0.44$ GRBs per year will have $T_{90}>30 \s$.

The realization that long GRBs can also emanate from binary mergers has profound implications on the follow-up program of future GW observations. First of all, if a GRB presents a spike and extended emission structure, follow-up is warranted. Here we propose that fast variations in the lightcurve may be a distinguishing feature of mergers.
It is more likely however that the variations have a continuous distribution and \grb is special even among the short GRBs with extended emission. Indeed we found that only one long GRB has comparably short MVT. The flux and fluence of \grb are both extreme among the \Fermi-GBM GRBs.

{\it Acknowledgements-}
UAH co-authors acknowledge NASA funding from cooperative agreement 80MSFC22M0004. NF is grateful to UNAM-DGAPA-PAPIIT for the funding provided by grant IN106521.
USRA co-authors acknowledge NASA funding from cooperative agreement 80MSFC17M0022.

\bibliographystyle{apj}

\begin{thebibliography}{}
\expandafter\ifx\csname natexlab\endcsname\relax\def\natexlab#1{#1}\fi

\bibitem[{{Abbott} {et~al.}(2017){Abbott}, {Abbott}, {Abbott}, {Acernese},
  {Ackley}, {Adams}, {Adams}, {Addesso}, {Adhikari}, {Adya}, \&
  et~al.}]{Abbott2017aGWGRB}
{Abbott}, B.~P., {Abbott}, R., {Abbott}, T.~D., {et~al.} 2017, \apjl, 848, L13

\bibitem[{Abbott {et~al.}(2021)}]{LIGOScientific:2020lst}
Abbott, R., {et~al.} 2021, Astrophys. J., 915, 86

\bibitem[{Abbott {et~al.}(2022)}]{LIGOScientific:2021iyk}
---. 2022, Astrophys. J., 928, 186

\bibitem[{{Ajello} {et~al.}(2019){Ajello}, {Arimoto}, {Axelsson}, {Baldini},
  {Barbiellini}, {Bastieri}, {Bellazzini}, {Bhat}, {Bissaldi}, \&
  {Blandford}}]{Ajello+19LATcat2}
{Ajello}, M., {Arimoto}, M., {Axelsson}, M., {et~al.} 2019, \apj, 878, 52

\bibitem[{{Ajello} {et~al.}(2020){Ajello}, {Arimoto}, {Axelsson}, {Baldini},
  {Barbiellini}, {Bastieri}, {Bellazzini}, {Berretta}, {Bissaldi}, {Blandford},
  {Bonino}, {Bottacini}, {Bregeon}, {Bruel}, {Buehler}, {Burns}, {Buson},
  {Cameron}, {Caputo}, {Caraveo}, {Cavazzuti}, {Chen}, {Chiaro}, {Ciprini},
  {Cohen-Tanugi}, {Costantin}, {Cutini}, {D'Ammando}, {DeKlotz}, {de la Torre
  Luque}, {de Palma}, {Desai}, {Di Lalla}, {Di Venere}, {Fana Dirirsa},
  {Fegan}, {Franckowiak}, {Fukazawa}, {Funk}, {Fusco}, {Gargano}, {Gasparrini},
  {Giglietto}, {Gill}, {Giordano}, {Giroletti}, {Granot}, {Green}, {Grenier},
  {Grondin}, {Guiriec}, {Hays}, {Horan}, {J{\'o}hannesson}, {Kocevski},
  {Kovac'evic'}, {Kuss}, {Larsson}, {Latronico}, {Lemoine-Goumard}, {Li},
  {Liodakis}, {Longo}, {Loparco}, {Lovellette}, {Lubrano}, {Maldera},
  {Malyshev}, {Manfreda}, {Mart{\'\i}-Devesa}, {Mazziotta}, {McEnery}, {Mereu},
  {Meyer}, {Michelson}, {Mitthumsiri}, {Mizuno}, {Monzani}, {Moretti},
  {Morselli}, {Moskalenko}, {Negro}, {Nuss}, {Omodei}, {Orienti}, {Orlando},
  {Palatiello}, {Paliya}, {Paneque}, {Pei}, {Persic}, {Pesce-Rollins},
  {Petrosian}, {Piron}, {Poon}, {Porter}, {Principe}, {Racusin}, {Rain{\`o}},
  {Rando}, {Rani}, {Razzano}, {Razzaque}, {Reimer}, {Reimer}, {Ryde}, {Saz
  Parkinson}, {Serini}, {Sgr{\`o}}, {Siskind}, {Spandre}, {Spinelli}, {Tajima},
  {Takagi}, {Takahashi}, {Tak}, {Thayer}, {Thompson}, {Torres}, {Troja},
  {Valverde}, {Van Klaveren}, {Wood}, {Yassine}, {Zaharijas}, {Mailyan},
  {Bhat}, {Briggs}, {Cleveland}, {Giles}, {Goldstein}, {Hui}, {Malacaria},
  {Preece}, {Roberts}, {Veres}, {Wilson-Hodge}, {Kienlin}, {Cenko}, {O'Brien},
  {Beardmore}, {Lien}, {Osborne}, {Tohuvavohu}, {D'Elia}, {D'A{\`\i}}, {Perri},
  {Gropp}, {Klingler}, {Capalbi}, {Tagliaferri}, {Stamatikos}, \& {De
  Pasquale}}]{Ajello+20190114c}
---. 2020, \apj, 890, 9

\bibitem[{{Band} {et~al.}(1993){Band}, {Matteson}, {Ford}, {Schaefer},
  {Palmer}, {Teegarden}, {Cline}, {Briggs}, {Paciesas}, {Pendleton}, {Fishman},
  {Kouveliotou}, {Meegan}, {Wilson}, \& {Lestrade}}]{Band+93}
{Band}, D., {Matteson}, J., {Ford}, L., {et~al.} 1993, \apj, 413, 281

\bibitem[{{Bhat} {et~al.}(2016){Bhat}, {Meegan}, {von Kienlin}, {Paciesas},
  {Briggs}, {Burgess}, {Burns}, {Chaplin}, {Cleveland}, {Collazzi},
  {Connaughton}, {Diekmann}, {Fitzpatrick}, {Gibby}, {Giles}, {Goldstein},
  {Greiner}, {Jenke}, {Kippen}, {Kouveliotou}, {Mailyan}, {McBreen}, {Pelassa},
  {Preece}, {Roberts}, {Sparke}, {Stanbro}, {Veres}, {Wilson-Hodge}, {Xiong},
  {Younes}, {Yu}, \& {Zhang}}]{Bhat+16cat}
{Bhat}, N.~P., {Meegan}, C.~A., {von Kienlin}, A., {et~al.} 2016, \apjs, 223,
  28

\bibitem[{{Bhat}(2013)}]{Bhat13tvar}
{Bhat}, P.~N. 2013, arXiv e-prints, arXiv:1307.7618

\bibitem[{{Bhat} {et~al.}(1992){Bhat}, {Fishman}, {Meegan}, {Wilson}, {Brock},
  \& {Paciesas}}]{Bhat+92shortvar}
{Bhat}, P.~N., {Fishman}, G.~J., {Meegan}, C.~A., {et~al.} 1992, \nat, 359, 217

\bibitem[{{Bhat} {et~al.}(2012){Bhat}, {Briggs}, {Connaughton}, {Kouveliotou},
  {van der Horst}, {Paciesas}, {Meegan}, {Bissaldi}, {Burgess}, {Chaplin},
  {Diehl}, {Fishman}, {Fitzpatrick}, {Foley}, {Gibby}, {Giles}, {Goldstein},
  {Greiner}, {Gruber}, {Guiriec}, {von Kienlin}, {Kippen}, {McBreen}, {Preece},
  {Rau}, {Tierney}, \& {Wilson-Hodge}}]{Bhat+12mvt}
{Bhat}, P.~N., {Briggs}, M.~S., {Connaughton}, V., {et~al.} 2012, \apj, 744,
  141

\bibitem[{{Bucciantini} {et~al.}(2012){Bucciantini}, {Metzger}, {Thompson}, \&
  {Quataert}}]{Bucciantini+12sgrbEE}
{Bucciantini}, N., {Metzger}, B.~D., {Thompson}, T.~A., \& {Quataert}, E. 2012,
  \mnras, 419, 1537

\bibitem[{{Burgess} {et~al.}(2009){Burgess}, {Goldstein}, \& {van der
  Horst}}]{2009GCN..9698....1B}
{Burgess}, J.~M., {Goldstein}, A., \& {van der Horst}, A.~J. 2009, GRB
  Coordinates Network, 9698, 1

\bibitem[{Burns {et~al.}(2021)Burns, Svinkin, Hurley, Wadiasingh, Negro,
  Younes, Hamburg, Ridnaia, Cook, Cenko, {et~al.}}]{burns2021identification}
Burns, E., Svinkin, D., Hurley, K., {et~al.} 2021, The Astrophysical Journal
  Letters, 907, L28

\bibitem[{{Connaughton}(2002)}]{connaughton02}
{Connaughton}, V. 2002, \apj, 567, 1028

\bibitem[{{D'Ai} {et~al.}(2021){D'Ai}, {Ambrosi}, {D'Elia}, {Gropp}, {Kennea},
  {Kuin}, {Lien}, {Marshall}, {Page}, {Palmer}, {Parsotan}, {Sbarufatti}, \&
  {Neil Gehrels Swift Observatory Team}}]{2021GCN.31202....1D}
{D'Ai}, A., {Ambrosi}, E., {D'Elia}, V., {et~al.} 2021, GRB Coordinates
  Network, 31202, 1

\bibitem[{{Dalessi} \& {Fermi GBM Team}(2023{\natexlab{a}})}]{Dalessi+23BTI}
{Dalessi}, S., \& {Fermi GBM Team}. 2023{\natexlab{a}}, GRB Coordinates
  Network, 33551, 1

\bibitem[{{Dalessi} \& {Fermi GBM
  Team}(2023{\natexlab{b}})}]{Dalessi+23230307a}
---. 2023{\natexlab{b}}, GRB Coordinates Network, 33407, 1

\bibitem[{{Duncan} \& {Thompson}(1992)}]{1992ApJ...392L...9D}
{Duncan}, R.~C., \& {Thompson}, C. 1992, \apjl, 392, L9

\bibitem[{{Fulton} {et~al.}(2023){Fulton}, {Smartt}, {Rhodes}, {Huber},
  {Villar}, {Moore}, {Srivastav}, {Schultz}, {Chambers}, {Izzo}, {Hjorth},
  {Chen}, {Nicholl}, {Foley}, {Rest}, {Smith}, {Young}, {Sim}, {Bright},
  {Zenati}, {de Boer}, {Bulger}, {Fairlamb}, {Gao}, {Lin}, {Lowe}, {Magnier},
  {Smith}, {Wainscoat}, {Coulter}, {Jones}, {Kilpatrick}, {McGill},
  {Ramirez-Ruiz}, {Lee}, {Narayan}, {Ramakrishnan}, {Ridden-Harper}, {Singh},
  {Wang}, {Kong}, {Ngeow}, {Pan}, {Yang}, {Davis}, {Piro}, {Rojas-Bravo},
  {Sommer}, \& {Yadavalli}}]{Fulton+23221009asupernova}
{Fulton}, M.~D., {Smartt}, S.~J., {Rhodes}, L., {et~al.} 2023, \apjl, 946, L22

\bibitem[{{Gehrels} {et~al.}(2006){Gehrels}, {Norris}, {Barthelmy}, {Granot},
  {Kaneko}, {Kouveliotou}, {Markwardt}, {M{\'e}sz{\'a}ros}, {Nakar}, {Nousek},
  {O'Brien}, {Page}, {Palmer}, {Parsons}, {Roming}, {Sakamoto}, {Sarazin},
  {Schady}, {Stamatikos}, \& {Woosley}}]{Gehrels+06strangegrb}
{Gehrels}, N., {Norris}, J.~P., {Barthelmy}, S.~D., {et~al.} 2006, \nat, 444,
  1044

\bibitem[{{Gendre} {et~al.}(2013){Gendre}, {Stratta}, {Atteia}, {Basa},
  {Bo{\"e}r}, {Coward}, {Cutini}, {D'Elia}, {Howell}, {Klotz}, \&
  {Piro}}]{Gendre2013}
{Gendre}, B., {Stratta}, G., {Atteia}, J.~L., {et~al.} 2013, \apj, 766, 30

\bibitem[{{Giblin} {et~al.}(1999){Giblin}, {van Paradijs}, {Kouveliotou},
  {Connaughton}, {Wijers}, {Briggs}, {Preece}, \&
  {Fishman}}]{Giblin+99afterglow}
{Giblin}, T.~W., {van Paradijs}, J., {Kouveliotou}, C., {et~al.} 1999, \apjl,
  524, L47

\bibitem[{{Goldstein} {et~al.}(2017){Goldstein}, {Veres}, {Burns}, {Briggs},
  {Hamburg}, {Kocevski}, {Wilson-Hodge}, {Preece}, {Poolakkil}, {Roberts},
  {Hui}, {Connaughton}, {Racusin}, {von Kienlin}, {Dal Canton}, {Christensen},
  {Littenberg}, {Siellez}, {Blackburn}, {Broida}, {Bissaldi}, {Cleveland},
  {Gibby}, {Giles}, {Kippen}, {McBreen}, {McEnery}, {Meegan}, {Paciesas}, \&
  {Stanbro}}]{Goldstein+17170817a}
{Goldstein}, A., {Veres}, P., {Burns}, E., {et~al.} 2017, \apjl, 848, L14

\bibitem[{{Golkhou} {et~al.}(2015{\natexlab{a}}){Golkhou}, {Butler}, \&
  {Littlejohns}}]{Golkhou+15tvar}
{Golkhou}, V.~Z., {Butler}, N.~R., \& {Littlejohns}, O.~M. 2015{\natexlab{a}},
  \apj, 811, 93

\bibitem[{{Golkhou} {et~al.}(2015{\natexlab{b}}){Golkhou}, {Butler}, \&
  {Littlejohns}}]{Golkhou+15variability}
---. 2015{\natexlab{b}}, \apj, 811, 93

\bibitem[{{Gompertz} {et~al.}(2023){Gompertz}, {Ravasio}, {Nicholl}, {Levan},
  {Metzger}, {Oates}, {Lamb}, {Fong}, {Malesani}, {Rastinejad}, {Tanvir},
  {Evans}, {Jonker}, {Page}, \& {Pe'er}}]{Gompertz+22grb211211a}
{Gompertz}, B.~P., {Ravasio}, M.~E., {Nicholl}, M., {et~al.} 2023, Nature
  Astronomy, 7, 67

\bibitem[{{Greiner} {et~al.}(2015){Greiner}, {Mazzali}, {Kann}, {Kr{\"u}hler},
  {Pian}, {Prentice}, {Olivares E.}, {Rossi}, {Klose}, {Taubenberger}, {Knust},
  {Afonso}, {Ashall}, {Bolmer}, {Delvaux}, {Diehl}, {Elliott}, {Filgas},
  {Fynbo}, {Graham}, {Guelbenzu}, {Kobayashi}, {Leloudas}, {Savaglio},
  {Schady}, {Schmidl}, {Schweyer}, {Sudilovsky}, {Tanga}, {Updike}, {van
  Eerten}, \& {Varela}}]{Greiner2015Natur}
{Greiner}, J., {Mazzali}, P.~A., {Kann}, D.~A., {et~al.} 2015, \nat, 523, 189

\bibitem[{{Guiriec} {et~al.}(2015){Guiriec}, {Mochkovitch}, {Piran}, {Daigne},
  {Kouveliotou}, {Racusin}, {Gehrels}, \&
  {McEnery}}]{Guiriec+15131014ashortbright}
{Guiriec}, S., {Mochkovitch}, R., {Piran}, T., {et~al.} 2015, \apj, 814, 10

\bibitem[{Harry \& Fairhurst(2011)}]{Harry:2010fr}
Harry, I.~W., \& Fairhurst, S. 2011, Phys. Rev. D, 83, 084002

\bibitem[{{Hjorth} {et~al.}(2003){Hjorth}, {Sollerman}, {M{\o}ller}, {Fynbo},
  {Woosley}, {Kouveliotou}, {Tanvir}, {Greiner}, {Andersen}, {Castro-Tirado},
  {Castro Cer{\'o}n}, {Fruchter}, {Gorosabel}, {Jakobsson}, {Kaper}, {Klose},
  {Masetti}, {Pedersen}, {Pedersen}, {Pian}, {Palazzi}, {Rhoads}, {Rol}, {van
  den Heuvel}, {Vreeswijk}, {Watson}, \& {Wijers}}]{hjorth03}
{Hjorth}, J., {Sollerman}, J., {M{\o}ller}, P., {et~al.} 2003, \nat, 423, 847

\bibitem[{{Ioka} \& {Nakamura}(2002)}]{Ioka+02lognormal}
{Ioka}, K., \& {Nakamura}, T. 2002, \apjl, 570, L21

\bibitem[{{Kaneko} {et~al.}(2015){Kaneko}, {Bostanc{\i}},
  {G{\"o}{\u{g}}{\"u}{\c{s}}}, \& {Lin}}]{Kaneko+15ee}
{Kaneko}, Y., {Bostanc{\i}}, Z.~F., {G{\"o}{\u{g}}{\"u}{\c{s}}}, E., \& {Lin},
  L. 2015, \mnras, 452, 824

\bibitem[{{Kann} {et~al.}(2011){Kann}, {Klose}, {Zhang}, {Covino}, {Butler},
  {Malesani}, {Nakar}, {Wilson}, {Antonelli}, {Chincarini}, {Cobb}, {D'Avanzo},
  {D'Elia}, {Della Valle}, {Ferrero}, {Fugazza}, {Gorosabel}, {Israel},
  {Mannucci}, {Piranomonte}, {Schulze}, {Stella}, {Tagliaferri}, \&
  {Wiersema}}]{kann2011}
{Kann}, D.~A., {Klose}, S., {Zhang}, B., {et~al.} 2011, \apj, 734, 96

\bibitem[{{Kann} {et~al.}(2018){Kann}, {Schady}, {Olivares}, {Klose}, {Rossi},
  {Perley}, {Zhang}, {Kr{\"u}hler}, {Greiner}, {Nicuesa Guelbenzu}, {Elliott},
  {Knust}, {Cano}, {Filgas}, {Pian}, {Mazzali}, {Fynbo}, {Leloudas}, {Afonso},
  {Delvaux}, {Graham}, {Rau}, {Schmidl}, {Schulze}, {Tanga}, {Updike}, \&
  {Varela}}]{Kann2018A&A}
{Kann}, D.~A., {Schady}, P., {Olivares}, E.~F., {et~al.} 2018, \aap, 617, A122

\bibitem[{{Kouveliotou} {et~al.}(1993){Kouveliotou}, {Meegan}, {Fishman},
  {Bhat}, {Briggs}, {Koshut}, {Paciesas}, \& {Pendleton}}]{kouveliotou93}
{Kouveliotou}, C., {Meegan}, C.~A., {Fishman}, G.~J., {et~al.} 1993, \apjl,
  413, L101

\bibitem[{{Lesage} {et~al.}(2023){Lesage}, {Veres}, {Briggs}, {Goldstein},
  {Kocevski}, {Burns}, {Wilson-Hodge}, {Bhat}, {Huppenkothen}, {Fryer},
  {Hamburg}, {Racusin}, {Bissaldi}, {Cleveland}, {Dalessi}, {Fletcher},
  {Giles}, {Hristov}, {Hui}, {Mailyan}, {Poolakkil}, {Roberts}, {von Kienlin},
  {Wood}, {Ajello}, {Arimoto}, {Baldini}, {Ballet}, {Baring}, {Bastieri},
  {Becerra Gonzalez}, {Bellazzini}, {Bissaldi}, {Blandford}, {Bonino}, {Bruel},
  {Buson}, {Cameron}, {Caputo}, {Caraveo}, {Cavazzuti}, {Chiaro}, {Cibrario},
  {Ciprini}, {Cristarella Orestano}, {Crnogorcevic}, {Cuoco}, {Cutini},
  {DAmmando}, {De Gaetano}, {Di Lalla}, {Di Venere}, {Dominguez}, {Fegan},
  {Ferrara}, {Fleischhack}, {Fukazawa}, {Funk}, {Fusco}, {Galanti}, {Gammaldi},
  {Gargano}, {Gasbarra}, {Gasparrini}, {Germani}, {Giacchino}, {Giglietto},
  {Gill}, {Giroletti}, {Granot}, {Green}, {Grenier}, {Guiriec}, {Gustafsson},
  {Hays}, {Hewitt}, {Horan}, {Hou}, {Kuss}, {Latronico}, {Laviron},
  {Lemoine-Goumard}, {Li}, {Liodakis}, {Longo}, {Loparco}, {Lorusso},
  {Lovellette}, {Lubrano}, {Maldera}, {Manfreda}, {Marti-Devesa}, {Mazziotta},
  {McEnery}, {Mereu}, {Meyer}, {Michelson}, {Mizuno}, {Monzani}, {Morselli},
  {Moskalenko}, {Negro}, {Nuss}, {Omodei}, {Orlando}, {Ormes}, {Paneque},
  {Panzarini}, {Persic}, {Pesce-Rollins}, {Pillera}, {Piron}, {Poon}, {Porter},
  {Principe}, {Raino}, {Rando}, {Rani}, {Razzano}, {Razzaque}, {Reimer},
  {Reimer}, {Ryde}, {Sanchez-Conde}, {Saz Parkinson}, {Scotton}, {Serini},
  {Sgro}, {Sharma}, {Siskind}, {Spandre}, {Spinelli}, {Tajima}, {Torres},
  {Valverde}, {Venters}, {Wadiasingh}, {Wood}, \&
  {Zaharijas}}]{Lesage+23221009a}
{Lesage}, S., {Veres}, P., {Briggs}, M.~S., {et~al.} 2023, arXiv e-prints,
  arXiv:2303.14172

\bibitem[{{Levan} {et~al.}(2014){Levan}, {Tanvir}, {Starling}, {Wiersema},
  {Page}, {Perley}, {Schulze}, {Wynn}, {Chornock}, {Hjorth}, {Cenko},
  {Fruchter}, {O'Brien}, {Brown}, {Tunnicliffe}, {Malesani}, {Jakobsson},
  {Watson}, {Berger}, {Bersier}, {Cobb}, {Covino}, {Cucchiara}, {de Ugarte
  Postigo}, {Fox}, {Gal-Yam}, {Goldoni}, {Gorosabel}, {Kaper}, {Kr{\"u}hler},
  {Karjalainen}, {Osborne}, {Pian}, {S{\'a}nchez-Ram{\'\i}rez}, {Schmidt},
  {Skillen}, {Tagliaferri}, {Th{\"o}ne}, {Vaduvescu}, {Wijers}, \&
  {Zauderer}}]{levan2014}
{Levan}, A.~J., {Tanvir}, N.~R., {Starling}, R.~L.~C., {et~al.} 2014, \apj,
  781, 13

\bibitem[{{Levan} {et~al.}(2023){Levan}, {Gompertz}, {Malesani}, {Tanvir},
  {Burns}, {Salvaterra}, {Ackley}, {Lamb}, {Fynbo}, {Schneider}, {Jakobsson},
  {Izzo}, {Fruchter}, {Watson}, {Kennedy}, {Hjorth}, {Pugliese},
  {Bhirombhakdi}, \& {Dhillon}}]{Levan+23230307aKN}
{Levan}, A.~J., {Gompertz}, B.~P., {Malesani}, D.~B., {et~al.} 2023, GRB
  Coordinates Network, 33569, 1

\bibitem[{{MacFadyen} \& {Woosley}(1999)}]{Macfadyen+99col}
{MacFadyen}, A.~I., \& {Woosley}, S.~E. 1999, \apj, 524, 262

\bibitem[{{MacLachlan} {et~al.}(2013){MacLachlan}, {Shenoy}, {Sonbas}, {Dhuga},
  {Cobb}, {Ukwatta}, {Morris}, {Eskandarian}, {Maximon}, \&
  {Parke}}]{MacLachlan+13tvar}
{MacLachlan}, G.~A., {Shenoy}, A., {Sonbas}, E., {et~al.} 2013, \mnras, 432,
  857

\bibitem[{{Malesani} {et~al.}(2021){Malesani}, {Fynbo}, {de Ugarte Postigo},
  {Izzo}, {Fu}, {Xu}, {Zhu}, {Tanvir}, \&
  {Djupvik}}]{Malesani+21211211aredshift}
{Malesani}, D.~B., {Fynbo}, J.~P.~U., {de Ugarte Postigo}, A., {et~al.} 2021,
  GRB Coordinates Network, 31221, 1

\bibitem[{{Mangan} {et~al.}(2021){Mangan}, {Dunwoody}, {Meegan}, \& {Fermi GBM
  Team}}]{2021GCN.31210....1M}
{Mangan}, J., {Dunwoody}, R., {Meegan}, C., \& {Fermi GBM Team}. 2021, GRB
  Coordinates Network, 31210, 1

\bibitem[{{Meegan} {et~al.}(2009){Meegan}, {Lichti}, {Bhat}, {Bissaldi},
  {Briggs}, {Connaughton}, {Diehl}, {Fishman}, {Greiner}, {Hoover}, {van der
  Horst}, {von Kienlin}, {Kippen}, {Kouveliotou}, {McBreen}, {Paciesas},
  {Preece}, {Steinle}, {Wallace}, {Wilson}, \& {Wilson-Hodge}}]{Meegan2009}
{Meegan}, C., {Lichti}, G., {Bhat}, P.~N., {et~al.} 2009, \apj, 702, 791

\bibitem[{{Mei} {et~al.}(2022){Mei}, {Banerjee}, {Oganesyan}, {Salafia},
  {Giarratana}, {Branchesi}, {D'Avanzo}, {Campana}, {Ghirlanda}, {Ronchini},
  {Shukla}, \& {Tiwari}}]{Mei+22211211aGeV}
{Mei}, A., {Banerjee}, B., {Oganesyan}, G., {et~al.} 2022, \nat, 612, 236

\bibitem[{{Minaev} {et~al.}(2021){Minaev}, {Pozanenko}, \& {GRB IKI
  FuN}}]{2021GCN.31230....1M}
{Minaev}, P., {Pozanenko}, A., \& {GRB IKI FuN}. 2021, GRB Coordinates Network,
  31230, 1

\bibitem[{{Morsony} {et~al.}(2010){Morsony}, {Lazzati}, \&
  {Begelman}}]{Morsony+10variab}
{Morsony}, B.~J., {Lazzati}, D., \& {Begelman}, M.~C. 2010, \apj, 723, 267

\bibitem[{{Narayan} \& {Kumar}(2009)}]{Narayan+09turb}
{Narayan}, R., \& {Kumar}, P. 2009, \mnras, L193+

\bibitem[{{Narayan} {et~al.}(1992){Narayan}, {Paczynski}, \&
  {Piran}}]{1992ApJ...395L..83N}
{Narayan}, R., {Paczynski}, B., \& {Piran}, T. 1992, \apjl, 395, L83

\bibitem[{{Norris} \& {Bonnell}(2006)}]{norris06}
{Norris}, J.~P., \& {Bonnell}, J.~T. 2006, \apj, 643, 266

\bibitem[{{Norris} {et~al.}(2005){Norris}, {Bonnell}, {Kazanas}, {Scargle},
  {Hakkila}, \& {Giblin}}]{Norris+05pulse}
{Norris}, J.~P., {Bonnell}, J.~T., {Kazanas}, D., {et~al.} 2005, \apj, 627, 324

\bibitem[{{Norris} {et~al.}(2010){Norris}, {Gehrels}, \&
  {Scargle}}]{Norris+10EE}
{Norris}, J.~P., {Gehrels}, N., \& {Scargle}, J.~D. 2010, \apj, 717, 411

\bibitem[{{Paciesas} {et~al.}(1999){Paciesas}, {Meegan}, {Pendleton}, {Briggs},
  {Kouveliotou}, {Koshut}, {Lestrade}, {McCollough}, {Brainerd}, {Hakkila},
  {Henze}, {Preece}, {Connaughton}, {Kippen}, {Mallozzi}, {Fishman},
  {Richardson}, \& {Sahi}}]{Paciesas+99cat}
{Paciesas}, W.~S., {Meegan}, C.~A., {Pendleton}, G.~N., {et~al.} 1999, \apjs,
  122, 465

\bibitem[{{Paczy{\'n}ski}(1998)}]{1998ApJ...494L..45P}
{Paczy{\'n}ski}, B. 1998, \apjl, 494, L45

\bibitem[{{Panaitescu} \& {Kumar}(2001)}]{panaitescu01}
{Panaitescu}, A., \& {Kumar}, P. 2001, \apjl, 560, L49

\bibitem[{{Piro} {et~al.}(2014){Piro}, {Troja}, {Gendre}, {Ghisellini},
  {Ricci}, {Bannister}, {Fiore}, {Kidd}, {Piranomonte}, \&
  {Wieringa}}]{Piro2014}
{Piro}, L., {Troja}, E., {Gendre}, B., {et~al.} 2014, \apjl, 790, L15

\bibitem[{{Planck Collaboration} {et~al.}(2020){Planck Collaboration},
  {Aghanim}, {Akrami}, {Ashdown}, {Aumont}, {Baccigalupi}, \& {et
  al.}}]{2020A&A...641A...6P}
{Planck Collaboration}, {Aghanim}, N., {Akrami}, Y., {et~al.} 2020, \aap, 641,
  A6

\bibitem[{{Poolakkil} {et~al.}(2021){Poolakkil}, {Preece}, {Fletcher},
  {Goldstein}, {Bhat}, {Bissaldi}, {Briggs}, {Burns}, {Cleveland}, {Giles},
  {Hui}, {Kocevski}, {Lesage}, {Mailyan}, {Malacaria}, {Paciesas}, {Roberts},
  {Veres}, {von Kienlin}, \& {Wilson-Hodge}}]{Poolakkil+21GBMspcat}
{Poolakkil}, S., {Preece}, R., {Fletcher}, C., {et~al.} 2021, \apj, 913, 60

\bibitem[{{Preece} {et~al.}(2014){Preece}, {Burgess}, {von Kienlin}, {Bhat},
  {Briggs}, {Byrne}, {Chaplin}, {Cleveland}, {Collazzi}, {Connaughton},
  {Diekmann}, {Fitzpatrick}, {Foley}, {Gibby}, {Giles}, {Goldstein}, {Greiner},
  {Gruber}, {Jenke}, {Kippen}, {Kouveliotou}, {McBreen}, {Meegan}, {Paciesas},
  {Pelassa}, {Tierney}, {van der Horst}, {Wilson-Hodge}, {Xiong}, {Younes},
  {Yu}, {Ackermann}, {Ajello}, {Axelsson}, {Baldini}, {Barbiellini}, {Baring},
  {Bastieri}, {Bellazzini}, {Bissaldi}, {Bonamente}, {Bregeon}, {Brigida},
  {Bruel}, {Buehler}, {Buson}, {Caliandro}, {Cameron}, {Caraveo}, {Cecchi},
  {Charles}, {Chekhtman}, {Chiang}, {Chiaro}, {Ciprini}, {Claus},
  {Cohen-Tanugi}, {Cominsky}, {Conrad}, {D'Ammando}, {de Angelis}, {de Palma},
  {Dermer}, {Desiante}, {Digel}, {Di Venere}, {Drell}, {Drlica-Wagner},
  {Favuzzi}, {Franckowiak}, {Fukazawa}, {Fusco}, {Gargano}, {Gehrels},
  {Germani}, {Giglietto}, {Giordano}, {Giroletti}, {Godfrey}, {Granot},
  {Grenier}, {Guiriec}, {Hadasch}, {Hanabata}, {Harding}, {Hayashida},
  {Iyyani}, {Jogler}, {J{\'o}hannesson}, {Kawano}, {Kn{\"o}dlseder},
  {Kocevski}, {Kuss}, {Lande}, {Larsson}, {Larsson}, {Latronico}, {Longo},
  {Loparco}, {Lovellette}, {Lubrano}, {Mayer}, {Mazziotta}, {Michelson},
  {Mizuno}, {Monzani}, {Moretti}, {Morselli}, {Murgia}, {Nemmen}, {Nuss},
  {Nymark}, {Ohno}, {Ohsugi}, {Okumura}, {Omodei}, {Orienti}, {Paneque},
  {Perkins}, {Pesce-Rollins}, {Piron}, {Pivato}, {Porter}, {Racusin},
  {Rain{\`o}}, {Rando}, {Razzano}, {Razzaque}, {Reimer}, {Reimer}, {Ritz},
  {Roth}, {Ryde}, {Sartori}, {Scargle}, {Schulz}, {Sgr{\`o}}, {Siskind},
  {Spandre}, {Spinelli}, {Suson}, {Tajima}, {Takahashi}, {Thayer}, {Thayer},
  {Tibaldo}, {Tinivella}, {Torres}, {Tosti}, {Troja}, {Usher}, {Vandenbroucke},
  {Vasileiou}, {Vianello}, {Vitale}, {Werner}, {Winer}, {Wood}, \&
  {Zhu}}]{Preece+14130427agbm}
{Preece}, R., {Burgess}, J.~M., {von Kienlin}, A., {et~al.} 2014, Science, 343,
  51

\bibitem[{{Rastinejad} {et~al.}(2022){Rastinejad}, {Gompertz}, {Levan}, {Fong},
  {Nicholl}, {Lamb}, {Malesani}, {Nugent}, {Oates}, {Tanvir}, {de Ugarte
  Postigo}, {Kilpatrick}, {Moore}, {Metzger}, {Ravasio}, {Rossi}, {Schroeder},
  {Jencson}, {Sand}, {Smith}, {Ag{\"u}{\'\i} Fern{\'a}ndez}, {Berger},
  {Blanchard}, {Chornock}, {Cobb}, {De Pasquale}, {Fynbo}, {Izzo}, {Kann},
  {Laskar}, {Marini}, {Paterson}, {Escorial}, {Sears}, \&
  {Th{\"o}ne}}]{Rastinejad+22grb211211a}
{Rastinejad}, J.~C., {Gompertz}, B.~P., {Levan}, A.~J., {et~al.} 2022, \nat,
  612, 223

\bibitem[{{Rees} \& {M\'esz\'aros}(1994)}]{Rees+94is}
{Rees}, M.~J., \& {M\'esz\'aros}, P. 1994, \apjl, 430, L93

\bibitem[{{Roberts} {et~al.}(2021){Roberts}, {Veres}, {Baring}, {Briggs},
  {Kouveliotou}, {Bissaldi}, {Younes}, {Chastain}, {DeLaunay}, {Huppenkothen},
  {Tohuvavohu}, {Bhat}, {G{\"o}{\v{g}}{\"u}{\c{s}}}, {van der Horst}, {Kennea},
  {Kocevski}, {Linford}, {Guiriec}, {Hamburg}, {Wilson-Hodge}, \&
  {Burns}}]{Roberts+21MGF}
{Roberts}, O.~J., {Veres}, P., {Baring}, M.~G., {et~al.} 2021, \nat, 589, 207

\bibitem[{{Rouco Escorial} {et~al.}(2021){Rouco Escorial}, {Fong}, {Veres},
  {Laskar}, {Lien}, {Paterson}, {Lally}, {Blanchard}, {Nugent}, {Tanvir},
  {Cornish}, {Berger}, {Burns}, {Cenko}, {Cobb}, {Cucchiara}, {Goldstein},
  {Margutti}, {Metzger}, {Milne}, {Levan}, {Nicholl}, \&
  {Smith}}]{Rouco+21180418}
{Rouco Escorial}, A., {Fong}, W., {Veres}, P., {et~al.} 2021, \apj, 912, 95

\bibitem[{{Rubtsov} {et~al.}(2012){Rubtsov}, {Pshirkov}, \&
  {Tinyakov}}]{2012MNRAS.421L..14R}
{Rubtsov}, G.~I., {Pshirkov}, M.~S., \& {Tinyakov}, P.~G. 2012, \mnras, 421,
  L14

\bibitem[{{Rybicki} \& {Lightman}(1979)}]{rybicki79}
{Rybicki}, G.~B., \& {Lightman}, A.~P. 1979, {Radiative processes in
  astrophysics} (New York, Wiley-Interscience, 1979.~393 p.)

\bibitem[{{Sari} \& {Piran}(1997{\natexlab{a}})}]{Sari+97intext}
{Sari}, R., \& {Piran}, T. 1997{\natexlab{a}}, \mnras, 287, 110

\bibitem[{{Sari} \& {Piran}(1997{\natexlab{b}})}]{Sari+97variab}
---. 1997{\natexlab{b}}, \apj, 485, 270

\bibitem[{{Sari} \& {Piran}(1999)}]{Sari+99optflash}
---. 1999, \apj, 520, 641

\bibitem[{{Sonbas} {et~al.}(2015){Sonbas}, {MacLachlan}, {Dhuga}, {Veres},
  {Shenoy}, \& {Ukwatta}}]{Sonbas+15LorentzMTS}
{Sonbas}, E., {MacLachlan}, G.~A., {Dhuga}, K.~S., {et~al.} 2015, \apj, 805, 86

\bibitem[{Sutton {et~al.}(2010)}]{Sutton:2009gi}
Sutton, P.~J., {et~al.} 2010, New J. Phys., 12, 053034

\bibitem[{{Tamura} {et~al.}(2021){Tamura}, {Yoshida}, {Sakamoto}, {Pal'Shin},
  {Sugita}, {Kawakubo}, {Yamaoka}, {Nakahira}, {Asaoka}, {Torii}, {Akaike},
  {Kobayashi}, {Shimizu}, {Cannady}, {Cherry}, {Ricciarini}, {Marrocchesi}, \&
  {Calet Collaboration}}]{2021GCN.31226....1T}
{Tamura}, T., {Yoshida}, A., {Sakamoto}, T., {et~al.} 2021, GRB Coordinates
  Network, 31226, 1

\bibitem[{{Tarnopolski}(2019)}]{Tarnopolski19skewed_distrib}
{Tarnopolski}, M. 2019, \apj, 870, 105

\bibitem[{{Thompson}(1994)}]{1994MNRAS.270..480T}
{Thompson}, C. 1994, \mnras, 270, 480

\bibitem[{{Troja} {et~al.}(2022){Troja}, {Fryer}, {O'Connor}, {Ryan},
  {Dichiara}, {Kumar}, {Ito}, {Gupta}, {Wollaeger}, {Norris}, {Kawai},
  {Butler}, {Aryan}, {Misra}, {Hosokawa}, {Murata}, {Niwano}, {Pandey},
  {Kutyrev}, {van Eerten}, {Chase}, {Hu}, {Caballero-Garcia}, \&
  {Castro-Tirado}}]{Troja+22grb211211a}
{Troja}, E., {Fryer}, C.~L., {O'Connor}, B., {et~al.} 2022, \nat, 612, 228

\bibitem[{{Usov}(1992)}]{1992Natur.357..472U}
{Usov}, V.~V. 1992, \nat, 357, 472

\bibitem[{{von Kienlin} {et~al.}(2020){von Kienlin}, {Meegan}, {Paciesas},
  {Bhat}, {Bissaldi}, {Briggs}, {Burns}, {Cleveland}, {Gibby}, {Giles},
  {Goldstein}, {Hamburg}, {Hui}, {Kocevski}, {Mailyan}, {Malacaria},
  {Poolakkil}, {Preece}, {Roberts}, {Veres}, \&
  {Wilson-Hodge}}]{vonKienlin+20GBM10yrcat}
{von Kienlin}, A., {Meegan}, C.~A., {Paciesas}, W.~S., {et~al.} 2020, arXiv
  e-prints, arXiv:2002.11460

\bibitem[{Was {et~al.}(2012)Was, Sutton, Jones, \& Leonor}]{Was:2012zq}
Was, M., Sutton, P.~J., Jones, G., \& Leonor, I. 2012, Phys. Rev. D, 86, 022003

\bibitem[{Williamson {et~al.}(2014)Williamson, Biwer, Fairhurst, Harry,
  Macdonald, Macleod, \& Predoi}]{Williamson:2014wma}
Williamson, A.~R., Biwer, C., Fairhurst, S., {et~al.} 2014, Phys. Rev. D, 90,
  122004

\bibitem[{{Wood} {et~al.}(2021){Wood}, {Meegan}, \& {Fermi GBM
  Team}}]{2021GCN.29788....1W}
{Wood}, J., {Meegan}, C., \& {Fermi GBM Team}. 2021, GRB Coordinates Network,
  29788, 1

\bibitem[{{Woosley}(1993)}]{1993ApJ...405..273W}
{Woosley}, S.~E. 1993, \apj, 405, 273

\bibitem[{{Woosley} \& {Bloom}(2006)}]{Woosley2006ARA&A}
{Woosley}, S.~E., \& {Bloom}, J.~S. 2006, \araa, 44, 507

\bibitem[{{Xiao} {et~al.}(2022){Xiao}, {Zhang}, {Zhu}, {Xiong}, {Gao}, {Xu},
  {Zhang}, {Peng}, {Li}, {Zhang}, {Lu}, {Lin}, {Liu}, {Zhang}, {Ge}, {Tuo},
  {Xue}, {Fu}, {Liu}, {Li}, {Wang}, {Zheng}, {Wang}, {Jiang}, {Li}, {Liu},
  {Cao}, {Cai}, {Yi}, {Zhao}, {Xie}, {Li}, {Luo}, {Liao}, {Song}, {Zhang},
  {Qu}, {Liu}, {Li}, {Xu}, \& {Li}}]{Xiao+22211211aperiod}
{Xiao}, S., {Zhang}, Y.-Q., {Zhu}, Z.-P., {et~al.} 2022, arXiv e-prints,
  arXiv:2205.02186

\bibitem[{{Yang} {et~al.}(2022){Yang}, {Zhang}, {Ai}, {Liu}, {Wang}, {Yang},
  {Yin}, {Li}, {L{\"u}}, \& {Zhang}}]{Yang+22211211a}
{Yang}, J., {Zhang}, B.~B., {Ai}, S.~K., {et~al.} 2022, arXiv e-prints,
  arXiv:2204.12771

\bibitem[{{Zhang} {et~al.}(2009){Zhang}, {Zhang}, {Virgili}, {Liang}, {Kann},
  {Wu}, {Proga}, {Lv}, {Toma}, {M{\'e}sz{\'a}ros}, {Burrows}, {Roming}, \&
  {Gehrels}}]{Zhang+09typeI}
{Zhang}, B., {Zhang}, B.-B., {Virgili}, F.~J., {et~al.} 2009, \apj, 703, 1696

\bibitem[{{Zhang} {et~al.}(2021){Zhang}, {Xiong}, {Li}, {Cai}, {Luo}, {Xiao},
  {Liu}, {Xue}, {Yi}, {Zheng}, {Li}, {Li}, {Liao}, {Song}, {Xiong}, {Liu},
  {Li}, {Li}, {Chang}, {Zhang}, {Zhang}, {Lu}, {Zou}, {Jin}, {Zhang}, {Li},
  {Lu}, {Song}, {Wu}, {Xu}, {Zhang}, \& {Insight-HXMT
  Team}}]{2021GCN.31236....1Z}
{Zhang}, Y.~Q., {Xiong}, S.~L., {Li}, X.~B., {et~al.} 2021, GRB Coordinates
  Network, 31236, 1

\end{thebibliography}

\end{document}